\documentclass[10pt]{revtex4}
\usepackage{amssymb}
\usepackage{latexsym}
\usepackage{epsfig}
\begin{document}
\title{Thermodynamic analysis of universes with the initial and final de-Sitter eras}
\author{H. Moradpour\footnote{h.moradpour@riaam.ac.ir}, M. T. Mohammadi Sabet\footnote{mohamadisabet@shirazu.ac.ir}, A. Ghasemi}
\address{Research Institute for Astronomy and Astrophysics of Maragha
(RIAAM), P.O. Box 55134-441, Maragha, Iran}

\begin{abstract}
Our aim is studying the thermodynamics of cosmological models
including initial and final de-Sitter eras. For this propose,
bearing Cai-Kim temperature in mind, we investigate the
thermodynamic properties of a dark energy candidate with variable
energy density, and show that the state parameter of this dark
energy candidate should obey the $\omega_D\neq-1$ constraint,
whiles there is no interaction between the fluids filled the
universe, and the universe is not in the de-Sitter eras.
Additionally, based on thermal fluctuation theory, we study the
possibility of inducing fluctuations to the entropy of the dark
energy candidate due to a mutual interaction between the cosmos
sectors. Therefore, we find a relation between the thermal
fluctuations and the mutual interaction between the cosmos
sectors, whiles the dark energy candidate has a varying energy
density. We point to models in which a gravitationally induced
particle production process leads to change the expansion eras,
whiles the corresponding pressure is considered as the cause of
current acceleration phase. We study its thermodynamics, and show
that such processes may also leave thermal fluctuations into the
system. We also find an expression between the thermal
fluctuations and the particle production rate. Finally, we use
Hayward-Kodama temperature to get a relation for the horizon
entropy in models including the gravitationally induced particle
production process. Our study shows that the first law of
thermodynamics is available on the apparent horizon whiles, the
gravitationally induced particle production process, as the dark
energy candidate, may add an additional term to the Bekenstein
limit of the horizon.
\end{abstract}
\pacs{98.80.-k, 95.36.+x}
\maketitle

\section{Introduction}
The universe expansion is modeled by the so-called FRW metric
\begin{eqnarray}\label{frw}
ds^{2}=dt^{2}-a^{2}\left( t\right) \left[ \frac{dr^{2}}{1-kr^{2}}%
+r^{2}d\Omega ^{2}\right],
\end{eqnarray}
describing a homogenous and isotropic cosmos. In this metric, $a(t)$
is the scale factor while $k=-1,0,1$ is the curvature parameter
corresponding to open, flat and closed universes respectively. WMAP
data indicates a flat universe \cite{roos}. Conformal form of this
metric can be used to model the inhomogeneity of cosmos due to the
structure abundance in scales smaller than $100$ Mpc \cite{rma}. The
universe expansion, in the standard cosmology ($\Lambda CDM$),
consists of four eras: ($\textmd{i}$) the primary inflation era
($a(t)\sim e^{Ht}$) exactly began after the big bang needed to
remove the horizon problem. ($\textmd{ii}$) the radiation dominated
era ($a(t)\sim t^{\frac{1}{2}}$) finished by leaving an attractive
trace which is now called either CMB or LSS. ($\textmd{iii}$) the
matter dominated era ($a(t)\sim t^{\frac{2}{3}}$) which has the
major part in the structure formation process. ($\textmd{iv}$)
finally, in agreement with the current phase of the expansion, the
universe is undergoing an accelerated phase ($a(t)\sim e^{Ht}$)
which is similar to the primary inflation era, where
$H=\frac{\dot{a}}{a}$ and dot are the Hubble parameter and
derivative with respect to time, respectively. In order to explain
both the primary inflation and the current phase of the expansion
leading to the coincidence and cosmological constant problems
\cite{roos} we need an abnormal matters. Thermodynamic
considerations show that the universe maintains this current phase
of the expansion forever leading to avoid the big crunch problem
\cite{pavon2}. We should note that this model ($\Lambda CDM$) is in
a very good agreement with the observations
\cite{roos,c1,c2,c3,c4,c5}. It should be noted that if one accepts
the thermodynamic predictions about the current stage of the
expansion \cite{pavon2}, then three questions including the big
bang, coincidence and cosmological constant problems remain
unsolved.

There are various models proposed to eliminate such weaknesses of
the standard cosmology by introducing either a new degree of freedom
or a new parameter leading to explanations for the generator of the
current acceleration phase (called dark energy (DE)) in the Einstein
relativity framework
\cite{de1,de2,de3,de4,de5,de6,ven,GGDE,jcap,mod}. Moreover, it is
shown that the DE candidates may add an additional term to the
Bekenstein entropy of the apparent and trapping horizons of the FRW
universe \cite{cana,cana1,em}. Recently, two models are introduced
in the literatures to solve the mentioned problems of the standard
cosmology whenever their physics are completely different from those
of previous works \cite{LBSV,LBC}. In these models, the universe
expansion is began from an unstable de-Sitter spacetime leading to
solve the horizon problem. In addition to this, the final stage of
the universe expansion is an eternal de-Sitter phase which is due to
the thermodynamic equilibrium conditions and leading to solve the
big crunch singularity problem \cite{pavon3,LBC1}. Because of
different physics behind these models, they become interesting for
further investigations \cite{l1,l2,l4,l5,l6,l7,l8}. As the main
difference of these two models, while a varying vacuum plays the
role of DE in the model proposed in \cite{LBSV}, the gravitationally
induced particle production process is the backbone of the second
proposal \cite{LBC,LBC1}. Moreover, it is shown that the first model
satisfies the Bekenstein limit of entropy \cite{em}.

It seems that the apparent horizon of the FRW universe, as a
marginally trapped surface which is located at \cite{sheyw1,sheyw2}
\begin{eqnarray}\label{ah}
\tilde{r}_A=\frac{1}{\sqrt{H^2+\frac{k}{a(t)^2}}},
\end{eqnarray}
can be considered as a causal boundary for this spacetime
\cite{Bak,Hay2,Hay22,wcom,wcom1,wcom2}. Surface gravity of the
apparent horizon is evaluated as
\begin{equation}\label{SG}
\kappa=\frac{1}{2\sqrt{-h}}\partial_{a}(\sqrt{-h}h^{ab}\partial_{b}\zeta),
\end{equation}
leading to
\begin{eqnarray}
\kappa=-\frac{H}{2\pi}(1+\frac{\dot{H}}{2H^2}),
\end{eqnarray}
and we get
\begin{eqnarray}\label{t}
T=\frac{\kappa}{2\pi}=-\frac{H}{2\pi}(1+\frac{\dot{H}}{2H^2}),
\end{eqnarray}
as the temperature on the apparent horizon \cite{sheyw1,sheyw2}.
Indeed, this temperature is called Hayward-Kodama temperature used
to show the consistency between the Friedmann equations and the
first law of thermodynamics \cite{Bak,hel,hel1,GSL1}. Moreover,
some authors define $T=\frac{|\kappa|}{2\pi}\simeq\frac{H}{2\pi}$,
called Cai-Kim temperature, in order to get the positive value for
temperature \cite{CaiKim}. Another motivation for
$T=\frac{H}{2\pi}$, signalling us that this temperature may be
considered as the temperature for fields confined to the apparent
horizon, can be found in ref.~\cite{Cai44}. Finally, it is useful
to note that these definitions of temperature could not attract a
common agreement \cite{hel,hel1,GSL1}.

Bearing the various definitions of temperature in mind, since the
apparent horizon can be considered as the causal boundary, some
authors have been shown that the validity of the first law of
thermodynamics on the apparent horizon leads to Friedmann
equations
\cite{shey2,Cai2,Cai3,CaiKim,Wang,Wangg,Wanggg,Cai4,GSL1,Bak,Cai44,hel,hel1}.
In addition, it seems that it is necessary to consider a DE
\cite{roos,rev1,rev2} or modifying the Einstein equations
\cite{meg} in order to be compatible with recent observations
imposing the $\dot{a}(t)\geq0$ and $\ddot{a}(t)\geq0$ conditions
on the scale factor \cite{Rie,Rie1,Rie2,Rie3}. These data are also
in agreement with the generalized second law of thermodynamics in
numerous models of DE providing an eternal thermodynamic
equilibrium state for the universe
\cite{pavon1,pavon2,pavon3,msrw,mr}. More studies on the
thermodynamics of final state of the cosmos can be found in ref.
\cite{noj,noj1}.

Moreover, observations admit an interaction between the dark
sectors of the cosmos \cite{ob1,ob2,ob3,ob4,ob5,ob6,ob7}.
Additionally, it seems that the mutual interaction between the
dark sectors of the universe may solve the coincidence problem
\cite{ob7,co1,co2,co3,co4,co5,co6,pavonz}. Considering thermal
fluctuation theory in mind \cite{landau}, authors have shown that
the entropy of event horizon is modified by a logarithmic
correction \cite{das}. In the cosmological setup, by using the
Cai-Kim temperature, it is shown that these fluctuations may be
interpreted as the result of a mutual interaction between the dark
components \cite{pavonz}. The generalization of this approach to
more cosmological models can be found in
\cite{shs0,shs00,shs,shs1,int3,em1}. Therefore, it seems that one
can find an expression for the mutual interaction between the dark
sectors of the cosmos by taking into account the thermal
fluctuations of the universe components. In addition, it is shown
that a mutual interaction between the DE candidate and the other
parts of cosmos may add an additional term to the Bekenstein
entropy of the apparent horizon \cite{em}.

Here, according to the foregoing discussion and by following the
approach considered in
refs.~\cite{pavonz,shs0,shs00,shs,shs1,int3,em1}, we focus on the
models proposed in Refs.~\cite{LBSV,LBC}, and try to find the
suitable thermodynamic interpretations for these models by using
the Cai-Kim temperature as well as the thermal fluctuation theory.
In fact, we try to get a thermodynamic interpretation for the
mutual interaction between the DE candidate and the other parts of
cosmos. We also use the Hayward-Kodama temperature, and show that
the gravitationally induced particle production process, as the DE
candidate, may add an additional term to the Bekenstein entropy of
the apparent horizon. The latter is in agreement with the previous
attempts which predict that the DE candidates may correct the
Bekenstein bound in the cosmological setups \cite{cana,cana1,em}.
For simplicity, we take $G=\hbar=c=1$ throughout this paper and we
restrict ourselves to the $k=0$ case. Moreover, dot denotes
derivative with respect to time.

The paper is organized as follows. In next section, we decompose the
universe sectors into a varying DE candidate and other parts with total
density $\rho$ and state parameter $\omega$ whiles, the
cosmos sectors do not interact with each other. Then, we study the
thermodynamics of the DE candidate in this model. In section $\textmd{III}$, we
consider the model introduced in~\cite{LBSV}, and give a
thermodynamical interpretation for this model using thermal
fluctuations theory. To do this, the model will be considered as a
model in which the cosmos sectors are interacting with each other.
In section $\textmd{IV}$, we focus on the model proposed in
Ref.~\cite{LBC}, and show that the gravitationally induced particle
production process leads to the thermal fluctuations and finally, we
get a thermodynamical description for this model. We also show that
the gravitationally induced particle production process may add an
additional term to the Bekenstein limit of the apparent horizon of
the flat FRW universe in section $\textmd{V}$. Last section is
devoted to summary and concluding remarks.
\section{Thermodynamical description of DE models with non-constant energy density}\label{VDE}
For the flat FRW universe supported a DE candidate, Friedmann
equations lead to
\begin{eqnarray}\label{fried1}
H^{2}=\frac{8\pi}{3}(\rho+\rho_{D}),
\end{eqnarray}
and
\begin{eqnarray}\label{fried2}
-2\frac{\ddot{a}}{a}-H^2=8\pi G(p+p_D)
\end{eqnarray}
where $\rho_D$ and $p_D$ are the density of DE and its corresponding
pressure, respectively. In addition, $\rho$ comprises other parts of
cosmos which may include radiation, pressureless matter, dark matter
and etc. $p$ is also the pressure corresponding to the density
$\rho$. Consider a DE candidate with profile density
\begin{eqnarray}\label{den}
\rho_D(H)=\frac{\Lambda(H)}{8\pi}=\alpha+\beta H^2+\gamma H^{2n},
\end{eqnarray}
which converges to that of the ghost dark energy model by substituting
$\alpha=\beta=0$ and $n=\frac{1}{2}$ \cite{ven}. Moreover, it covers the
profile density of generalized ghost dark energy model by inserting $\alpha=0$
together with $n=\frac{1}{2}$ \cite{GGDE}. The cosmological constant model, as
the trivial limit, is obtainable by inserting $\beta=\gamma=0$ \cite{roos}. More
models in which authors used a dynamic DE model to explain the current expanding phase
can be achieved by choosing proper values for $n$, $\alpha$, $\beta$ and $\gamma$ \cite{LBSV,m1,m2,m3,m4,m5,m6,m7}.
It is also shown that this profile density may add an
additional term to the Bekenstein entropy of the trapping and
apparent horizons of the FRW universe \cite{cana,cana1,em}.
Using this equation, one gets
\begin{eqnarray}\label{dmden}
\rho=H^2(\frac{3}{8\pi}-\beta)-\alpha-\gamma H^{2n}.
\end{eqnarray}
For the total energy momentum tensor, the energy momentum
conservation law implies
\begin{eqnarray}\label{emc}
\dot{\rho}+\dot{\rho}_{D}+3H(\rho(1+\omega)+\rho_D(1+3\omega_D))=0,
\end{eqnarray}
where dot denotes derivative with respect to time. This equation can
be decomposed into
\begin{eqnarray}\label{dmc1}
\dot{\rho}+3H\rho(1+\omega)=0,
\end{eqnarray}
and
\begin{eqnarray}\label{dec1}
\dot{\rho}_{D}+3H\rho_{D}(1+\omega_{D})=0,
\end{eqnarray}
where $\omega_i=\frac{p_i}{\rho_i}$ and $p_i$ are the state
parameter and the pressure of the i$^{\textmd{th}}$ sector,
respectively. Such well-advised decompositions are valid whenever
there is no mutual interaction between the dark sectors. In such a
situation, from Eqs.~(\ref{den}) and~(\ref{dec1}), clearly we have
$\dot{\rho}_D\neq0$ yielding $\omega_{D}\neq-1$. For models
introduced in Ref. \cite{LBSV}, the energy profile density of
varying vacuum, as the DE candidate, is also given by~(\ref{den}),
whiles $\omega_D=-1$, and satisfies the thermodynamic equilibrium
conditions in current acceleration phase of the universe expansion
\cite{pavon3}. It is shown that such models may avoid the big bang
and big crunch singularities as well as the horizon problem, and
can also provide a complete description for the history expansion
of the universe \cite{LBSV}. We should note here that since the
state parameter of the DE candidate satisfies the $\omega_D=-1$
constraint in model proposed by Lima et al.~\cite{LBSV}, the
decomposition of~(\ref{emc}) into~(\ref{dmc1}) and~(\ref{dec1}) is
possible if $\dot{\rho}=0$ leading to $\dot{H}=0$ because
$\dot{\rho}=\frac{d\rho}{dH}\dot{H}$. Briefly, since model
proposed by Lima et al.~\cite{LBSV} satisfies the thermodynamic
equilibrium conditions in the current de-Sitter accelerating phase
($\dot{H}=0$) \cite{pavon3}, the decomposition of~(\ref{emc}) into
Eqs.~(\ref{dmc1}) and~(\ref{dec1}) as the result of marginally
thermodynamic equilibrium states (de-Sitter spacetimes with
$\dot{H}=0$) of the model is reasonable.

Derivation from
Eq.~(\ref{fried1}) with respect to $t$ and using~(\ref{emc}), leads
to (Raychaudhuri equation)
\begin{eqnarray}\label{hdot1}
\dot{H}=\frac{dH}{dt}=-4\pi[\rho(1+\omega)+\rho_D(1+\omega_D)].
\end{eqnarray}
Now, if we define $\rho_c\equiv3\frac{H^2}{8\pi}$ and use
Eqs.~(\ref{fried1}),~(\ref{den}) and~(\ref{emc}) we obtain
\begin{eqnarray}\label{hdot11}
\dot{H}=-4\pi[(\omega_D-\omega)(\alpha+\beta H^2+\gamma
H^{2n})+\frac{3H^2}{8\pi}(1+\omega))].
\end{eqnarray}
Since we have only used Eqs.~(\ref{fried1}$-$\ref{emc}) in order to
derive Eqs.~(\ref{hdot1}) and~(\ref{hdot11}), we should note that
these equations are independent of the validity of Eqs.~(\ref{dmc1})
and~(\ref{dec1}), and thus the probable mutual interaction between
the dark sectors. Indeed, the validity of equation~(\ref{hdot1}),
and therefore~(\ref{hdot11}), is due to the Bianchi identity or the
conservation of the total energy momentum tensor~(\ref{emc}).
Finally, by inserting $\rho_c$ into the Friedmann equation we get
\begin{eqnarray}\label{friedman11}
1=\Omega_D+\Omega,
\end{eqnarray}
where $\Omega_i=\frac{\rho_i}{\rho_c}$ is the fractional energy
density of the i$^{\textmd{th}}$ component of the cosmos. For the DE
candidate confined to the flat FRW universe enclosed by the apparent
horizon~(\ref{ah}), the Gibb's law implies
\begin{eqnarray}\label{flt}
TdS_{D}=dE_{D}+p_{D}dV.
\end{eqnarray}
In this equation, $S_D$ is associated entropy to the DE while
$V=\frac{4\pi }{3}\tilde{r}_{A}^{3}$ and $E_D=\rho_{D}V$ are the
volume of the flat FRW universe and the energy of DE,
respectively. Additionally, thermodynamic equilibrium condition
implies that $T$ (the temperature of DE) should has the same value
as the temperature of the apparent horizon. Moreover, since it is
unreasonable to have a fluid with negative temperature, meaning
that $T>0$, the Cai-Kim temperature may be a good option for the
temperature of the DE candidate
\cite{Wanggg,sheyw2,Cai2,wcom2,CaiKim,GSL1,pavonz,shs0,shs00,shs,shs1,int3,em1}
\begin{eqnarray}\label{temp}
T=\frac{H}{2\pi }=\frac{1}{2\pi \tilde{r}_{A}},
\end{eqnarray}
where $\tilde{r}_{A}$ is the apparent horizon radius of the flat FRW
universe~(\ref{ah})
\begin{eqnarray}\label{ah2}
\tilde{r}_{A}=\frac{1}{H}.
\end{eqnarray}
Therefore, for the volume and energy differentials we reach
\begin{eqnarray}\label{vo}
dV=4\pi(\tilde{r}_{A})^{2}d\tilde{r}_{A}=-4\pi H^{-4}dH,
\end{eqnarray}
and
\begin{eqnarray}\label{ener}
dE_{D}=\rho_D dV+Vd\rho_D,
\end{eqnarray}
leading to
\begin{eqnarray}\label{f1}
dS_D=2\pi \tilde{r}_{A}
(\rho_D(1+\omega_D)4\pi(\tilde{r}_{A})^{2}d\tilde{r}_{A}
+\frac{4\pi}{3}(\tilde{r}_{A})^{3}d\rho_D),
\end{eqnarray}
where we have used $p_{D}=\rho_{D}\omega_{D}$. In addition, since
\begin{eqnarray}\label{rho1}
\dot{\rho}_D=\frac{d\rho_D}{d\tilde{r}_{A}}\frac{d\tilde{r}_{A}}{dH}\frac{dH}{dt},
\end{eqnarray}
we find
\begin{eqnarray}\label{rho11}
\frac{d\rho_D}{d\tilde{r}_{A}}=-\frac{3\rho_D(1+\omega_D)}{4\pi
\tilde{r}_{A}^3 (\rho(1+\omega)+\rho_D(1+\omega_D))},
\end{eqnarray}
where we have used Eqs.~(\ref{dec1}),~(\ref{hdot1}) and~(\ref{ah2}).
By combining this equation with~(\ref{f1}) and after some algebra,
we obtain
\begin{eqnarray}\label{entropy1}
dS_{D}^0=8\pi^2\rho_D^0(1+\omega_D^0)(\tilde{r}_{A}^0)^3
(1-\frac{1}{4\pi(\tilde{r}_{A}^0)^2(\rho^0(1+\omega^0)+\rho_D^0
(1+\omega_D^0))})d\tilde{r}_{A}^0.
\end{eqnarray}
We must note that the superscript ($0$) is used to indicate that
this result is valid whenever there is no interaction between the
cosmos sectors of the universe and therefore,
$\tilde{r}_{A}^0=\frac{1}{H_0}$. Since we did not use
Eq.~(\ref{den}) to obtain this equation, this result
(Eq.~(\ref{entropy1})) is also valid in every cosmological model
with the same Friedmann equation as Eq.~(\ref{fried1}). Using
Eqs.~(\ref{den}),~(\ref{hdot11}) and~(\ref{ah2}) the above
equation can be rewritten as follows
\begin{eqnarray}\label{entropy11}
dS_{D}^0=-\frac{8\pi^2(\alpha+\beta H_0^2+\gamma
H_0^{2n})(1+\omega_D^0)}{H_0^5}
(1-\frac{H_0^2}{4\pi[(\omega_D^0-\omega^0)(\alpha+\beta H_0^2+
\gamma H_0^{2n})+\frac{3H_0^2}{8\pi}(1+\omega^0)]})dH_0.
\end{eqnarray}
Now, using Eqs.~(\ref{den}) and~(\ref{friedman11}) to get
\begin{eqnarray}\label{entropy111}
\frac{dS_{D}^0}{dH_0}=\frac{3\pi\Omega_D^0(1+\omega_D^0)}{H_0^3}
(\frac{2}{3[(\omega_D^0-\omega^0)\Omega_D^0+(1+\omega^0)]}-1),
\end{eqnarray}
where the superscript/subscript ($0$) implies non-interacting
case. Therefore, we find an expression for the entropy of the DE
candidate when its energy density profile is varying with time as
introduced in Eq.~(\ref{den}). It is useful to be noted that, in
the $\omega^0_D=-1$ limit, $dS^0_D=0$ is obtainable meaning that
$\dot{\rho}_D=0$ which can be considered as either the
cosmological constant model of DE \cite{roos}, or the marginally
thermodynamic equilibrium states of model proposed by Lima et
al.~\cite{LBSV}. In fact, from~(\ref{entropy111}) it is apparent
that, whiles $\omega_D^0=-1$,
$\frac{dS_{D}^0}{dH_0}=\frac{d^2S_{D}^0}{dH^2_0}=0$ is available,
meaning that the thermodynamic equilibrium conditions are
marginally satisfied in these eras \cite{CALEN}. The latter
signals us that the initial and final de-Sitter spacetimes of
models proposed by Lima et al.~\cite{LBSV} are marginally
thermodynamic equilibrium states. In the next section, we show
that whenever a mutual interaction between the cosmos sectors
moves the cosmos between these two marginally thermodynamic
equilibrium states \cite{LBSV}, it leaves a thermal fluctuations
into the system.
\section{Thermodynamic description of the interacting DE models with non-constant energy density}\label{Int}
Here, we study the thermodynamics of a universe filled by a varying
vacuum, as the DE candidate, together with another source of energy
with density $\rho$ interacting with each other whenever the energy
density of the DE candidate is the same as the previous
section~(\ref{den}). In this manner, decomposition of the
energy-momentum conservation law~(\ref{emc}) implies
\begin{eqnarray}\label{emcigde}
\dot{\rho}+3H\rho(1+\omega)=-Q,
\end{eqnarray}
and
\begin{eqnarray}\label{emcigde2}
\dot{\rho}_{D}+3H\rho_{D}(1+\omega_{D})=Q,
\end{eqnarray}
where $Q$ is the mutual interaction between the cosmos sectors.
Clearly, $\omega_D=-1$ is accessible in this model while
$\dot{\rho}_D\neq0$. Some authors have considered this possibility
($\omega_D=-1$), and showed that the model leads to the compatible
outcomes with observational data and the thermodynamic equilibrium
conditions \cite{pavon3,LBSV}. In addition, it is also shown that,
independent of the profile density of the DE candidate, the
apparent horizon satisfies the Bekenstein entropy, in the
interacting models with $\omega_D=-1$ \cite{em}. It seems that the
mutual interaction between the dark sectors induce some
fluctuations into the thermodynamic properties of systems which
can be investigated by thermal fluctuations theory
\cite{das,pavonz,shs0,shs00,shs1,shs,int3,em1}. Due to these
fluctuations, the entropy is changed to $S_D$ and it can be
expanded as \cite{landau,das,pavonz,shs0,shs00,shs1,shs,int3,em1}
\begin{eqnarray}\label{totalentropy}
S_D=S_D^0+S_D^1+S^2_D
\end{eqnarray}
In this equation, $S_D^0$ is the entropy of the DE candidate when
there is no mutual interaction between the cosmos sectors of the
universe and $S_D^1=-\frac{1}{2}\ln CT_0^2$ is logarithmic
correction to the entropy where $C=T_0\frac{dS_D^0}{dT_0}$ is the
dimensionless heat capacity. $S^2_D$ also concerns higher order
terms. It is also useful to mention that this analysis is valid for
all thermodynamical systems \cite{das}. Combining
Eqs.~(\ref{entropy1}) and~(\ref{temp}), we get
\begin{eqnarray}\label{C1}
C=T_0\frac{dS_D^0}{dT_0}=-\frac{\rho_D^0(1+\omega_D^0)}{T_0^4}
(1-\frac{\pi T_0^2}{\rho^0(1+\omega^0)+\rho_D^0(1+\omega_D^0)}).
\end{eqnarray}
From Eq.~(\ref{temp}) we have also
\begin{eqnarray}\label{C11}
C=T_0\frac{dS_D^0}{dT_0}=H_0\frac{dS_D^0}{dH_0},
\end{eqnarray}
leading to
\begin{eqnarray}\label{C111}
C=-\frac{8\pi^2(\alpha+\beta H_0^2+\gamma
H_0^{2n})(1+\omega_D^0)}{H_0^4}
(1-\frac{H_0^2}{4\pi[(\omega_D^0-\omega^0) (\alpha+\beta
H_0^2+\gamma H_0^{2n})+\frac{3H_0^2}{8\pi}(1+\omega^0)]}),
\end{eqnarray}
where we have used Eq.~(\ref{entropy11}) in order to evaluate the
heat capacity of the DE candidate with profile density~(\ref{den}).
Therefore, one gets
\begin{eqnarray}\label{S1d}
S^1_D=-\frac{1}{2}\ln([\frac{32\pi^4(\alpha+\beta H_0^2+\gamma
H_0^{2n})
(1+\omega_D^0)}{H_0^2}(\frac{H_0^2}{4\pi[(\omega_D^0-\omega^0)(\alpha+\beta
H_0^2+\gamma H_0^{2n})+\frac{3H_0^2}{8\pi}(1+\omega^0)]}-1)]).
\end{eqnarray}
In deriving this equation we used again Eqs.~(\ref{temp})
and~(\ref{ah2}) along as~(\ref{C111}). It is a matter of calculation
to show that
\begin{eqnarray}\label{entropys1}
&&\frac{dS_D^1}{dH_0}=-\frac{1}{2}(\frac{2\beta H_0+2n\gamma
H_0^{2n-1}} {\alpha+\beta H_0^2+\gamma
H_0^{2n}}+\frac{2H_0-4\pi[(\omega_D^0-\omega^0)(2\beta H_0+2n\gamma
H_0^{2n-1})+\frac{3H_0}{4\pi}(1+\omega^0)]}{H_0^2-4\pi[(\omega_D^0-\omega^0)(\alpha+\beta
H_0^2+\gamma
H_0^{2n})+\frac{3H_0^2}{8\pi}(1+\omega^0)]}\nonumber \\
&-&\frac{[(\omega_D^0-\omega^0)(2\beta H_0+2n\gamma
H_0^{2n-1})+\frac{3H_0}{4\pi}(1+\omega^0)]}{[(\omega_D^0-\omega^0)(\alpha+\beta
H_0^2+\gamma
H_0^{2n})+\frac{3H_0^2}{8\pi}(1+\omega^0)]}-\frac{2}{H_0}),
\end{eqnarray}
which can be simplified, by using Eq.~(\ref{den}), as
\begin{eqnarray}\label{entropys11}
\frac{dS_D^1}{dH_0}=-\frac{1}{2}(\frac{(\rho_D^0)^{\prime}}{\rho_D^0}
+\frac{2H_0-4\pi[(\omega_D^0-\omega^0)(\rho_D^0)^{\prime}
+\frac{3H_0}{4\pi}(1+\omega^0)]}{H_0^2-4\pi[(\omega_D^0-\omega^0)
\rho_D^0+\frac{3H_0^2}{8\pi}(1+\omega^0)]}-\frac{[(\omega_D^0-\omega^0)(\rho_D^0)^{\prime}+\frac{3H_0}{4\pi}(1+\omega^0)]}
{[(\omega_D^0-\omega^0)\rho_D^0
+\frac{3H_0^2}{8\pi}(1+\omega^0)]}-\frac{2}{H_0}),
\end{eqnarray}
where the prime stands for the derivative with respect to $H_0$.
Again, we should note that the subscript/superscript ($0$) is used
to emphasize the non-interacting parameters. By using Friedmann
equation~(\ref{fried1}) along as Eqs.~(\ref{ah2})
and~(\ref{emcigde2}), it is easy to show that
\begin{eqnarray}\label{drho2}
\frac{d\rho_D}{d\tilde{r}_A}=\frac{Q-3H\rho_D(1+\omega_D)}{4\pi\tilde{r}_A^2
(\rho(1+\omega)+\rho_D(1+\omega_D))}.
\end{eqnarray}
By following the recipe of previous section we get
\begin{eqnarray}\label{entropy2}
dS_{D}=8\pi^2\rho_D(1+\omega_D)(\tilde{r}_{A})^3
(1+\frac{Q-3H\rho_D(1+\omega_D)}{12\pi\tilde{r}_{A}\rho_D(1+\omega_D)
(\rho(1+\omega)+\rho_D(1+\omega_D))})d\tilde{r}_{A}.
\end{eqnarray}
Using Eq.~(\ref{ah2}), we find
\begin{eqnarray}\label{entropy2n}
\frac{dS_D}{dH}=\frac{8\pi^2\rho_D(1+\omega_D)}{H^5}
(\frac{3H^2\rho_D(1+\omega_D)-QH}{12\pi \rho_D(1+\omega_D)
(\rho(1+\omega)+\rho_D(1+\omega_D))}-1).
\end{eqnarray}
As a desired result, the results of previous section are obtainable
by substituting $Q=0$. Bearing the model investigated in
\cite{LBSV,pavon3} in mind, where $\omega_D=-1$, and use
Eq.~(\ref{dmden}) to obtain
\begin{eqnarray}\label{entropy22}
\frac{dS_D}{dH}=-\frac{2\pi Q}{3(1+\omega)H^4(\alpha+\gamma
H^{2n}-H^2 (\frac{3}{8\pi}-\beta))}.
\end{eqnarray}
Comparing this equation together with~(\ref{entropy111}), we can
conclude that the mutual interaction may change the entropy of the
varying vacuum leading to separation from the marginally
thermodynamic equilibrium situation ($dS^0_D=d^2S^0_D=0$). Since
$Q=\dot{\rho}_D$ in this model, by using~(\ref{den})
and~(\ref{hdot11}), we get
\begin{eqnarray}\label{rhodotinter}
Q=4\pi\rho_D^{\prime}(\omega+1)(\rho_D-\frac{3H^2}{8\pi}).
\end{eqnarray}
Considering the definition of the fractional energy density, this
equation can written as
\begin{eqnarray}\label{rhodotinter2}
Q=\frac{3}{2}\rho_D^{\prime}(\omega+1)H^2(\Omega_D-1),
\end{eqnarray}
leading to
\begin{eqnarray}\label{ss}
\frac{dS_D}{dH}=-\frac{8\pi^2 \rho_{D}^{\prime}}{3H^4},
\end{eqnarray}
where again, prime stands for the derivative with respect to $H$.
We should note that the results of either considering cosmological
constant model of the DE or the initial and ultimate thermodynamic
equilibrium states of Lima et al. model~\cite{LBSV} (or briefly,
$dS_D=0$) are obtainable by inserting $\omega_D=-1$ and $Q=0$
simultaneously. Since Eq.~(\ref{totalentropy}) implies
\begin{eqnarray}\label{totalentropy1}
\frac{dS_D}{dH}=(\frac{dS^0_D}{dH_0}+\frac{dS^1_D}{dH_0}+\frac{dS^2_D}{dH_0})
\frac{dH_0}{dH},
\end{eqnarray}
we use Eq.~(\ref{entropy22}) to get
\begin{eqnarray}\label{Q1}
\frac{dS^2_D}{dH_0}=\frac{16\pi^2Q\dot{H}}{9(1+\omega_m)H^6(1-\Omega_D)\dot{H}_0}
-\frac{dS^0_D}{dH_0}-\frac{dS^1_D}{dH_0},
\end{eqnarray}
where, $\frac{dS^0_D}{dH_0}$ and $\frac{dS^1_D}{dH_0}$ are
introduced in Eqs.~(\ref{entropy111}) and~(\ref{entropys11}),
respectively. We have also used~(\ref{friedman11}) and
$\frac{dH}{dH_0}=\frac{\dot{H}}{\dot{H}_0}$ to derive this
equation, while $\dot{H}$ and $\dot{H}_0$ are evaluated by using
Eq.~(\ref{hdot11}) for the interacting and non-interacting cases,
respectively. Therefore, we find an expression for the thermal
fluctuations which are due to the interaction $Q=\dot{\rho}_D$.
Since it seems that the $S^2_D$ term has insignificant effects
with respect to the $S^1_D$ term in the gravitational and
cosmological setups
\cite{das,pavonz,shs00,shs0,shs,shs1,int3,em1}, we disregard it
and inserting Eq.~(\ref{ss}) into~(\ref{totalentropy1}), to get
\begin{eqnarray}\label{H}
\frac{\rho_D^{\prime}}{H^4}dH=-
\frac{3(dS^0_D+dS^1_D)}{8\pi^2}.
\end{eqnarray}
Now, using~(\ref{den}), we reach
\begin{eqnarray}\label{H1}
\frac{\beta}{H^2}+\frac{\gamma}{\frac{4}{2n}-1}H^{2(n-2)}=\frac{3}{8\pi^2}
(S^0_D+S^1_D+C),
\end{eqnarray}
where $C$ is an integration constant. Additionally, due to this fact
that entropy is not an absolute quantity, $C$ can be set to zero
without lose of generality. $S^1_D$ is evaluated in Eq.~(\ref{S1d}),
while $S^0_D$ can be obtained by integrating from
Eq.~(\ref{entropy11}). This is a relation for the Hubble parameter,
up to the first order fluctuations, due to the interaction $Q$.
Loosely speaking, based on the first order terms of the thermal
fluctuations which are due to the interaction between the DE
candidate and the other parts of cosmos, we find the mutual relation
between $H$ and $H_0$ in the model proposed in Ref.~\cite{LBSV}. It
should be noted that Eqs.~(\ref{Q1}-\ref{H1}) are only valid when
$\omega_D=-1$, $Q=\dot{\rho}_D$ while $\rho_D$ is explained by
Eq.~(\ref{den}). One can also find a general relation between an
un-known $Q$ and thermal fluctuations in the models with arbitrary
energy density for the DE, by using
Eqs.~(\ref{entropy1}),~(\ref{C1}),~(\ref{entropy2n})
and~(\ref{totalentropy1}).
\section{thermodynamical description of gravitationally particle production induced process }
Here, we focus on the LBC model proposed in Ref. \cite{LBC,LBC1}. In
this model, like the previous model \cite{LBSV}, the universe
expansion is began from an unstable initial de-Sitter spacetime and
follows the radiation and matter dominated era in continue. Finally,
the universe expansion will reach to a perpetual de-Sitter phase
which is in agreement with the thermodynamical equilibrium condition
\cite{pavon3,LBC1}. In this model, there is a particle production
due to the gravitational effects leading to an additional pressure
to the Friedman equations as:
\begin{eqnarray}\label{flbc1}
H^2=\frac{8\pi}{3}\rho,
\end{eqnarray}
and
\begin{eqnarray}\label{flbc}
-2\frac{\ddot{a}}{a}-H^2=8\pi(p+p_C),
\end{eqnarray}
where we considered a flat background \cite{LBC,LBC1}. $\rho$ is the
energy density of the dominated prefect fluid, such as the
radiation, confined to the apparent horizon of the FRW universe
whiles, $p$ is the corresponding pressure. The additional pressure
($p_C$) plays the key role in the change of the expansion phase, and
depends on the particle production rates. In a simple approach,
considering an adiabatic process for the particle creation, the
entropy per particle is constant whenever the total entropy is not
\cite{pp1}. Therefore, one reachs \cite{pp1,pp2,pp3,pp4,pp5,pp6}
\begin{eqnarray}\label{pre}
p_C=-\frac{\rho(1+\omega)\Gamma}{3H},
\end{eqnarray}
whiles $\Gamma$ is the particle production rate with dimension of
(time)$^{-1}$ \cite{LBC}. the Energy-momentum conservation law
implies
\begin{eqnarray}\label{emc2}
\dot{\rho}+3H\rho(1+\omega)=-Q,
\end{eqnarray}
where $Q=3Hp_C$ \cite{LBC,LBC1}. In the absence of this particle
production ($Q=0$), the entropy of the fluid with density $\rho$,
pressure $p$ and the state parameter $\omega$ reads as
\begin{eqnarray}\label{en1}
dS^0=-\frac{\pi}{H_0^3}(1+3\omega^0)dH_0,
\end{eqnarray}
where we have followed the recipe of section $\textmd{II}$. Indeed,
one can use Eqs.~(\ref{entropy1}) and~(\ref{ah2}), substitute
$\rho^0=0$, and finally replace $\rho_D$ with $\rho$ together with
using~(\ref{flbc1}) to get this equation. We should note that since
for the cosmological constant model $\dot{\rho}=0$ leading to
$dH_0=0$, this equation covers the result of considering the
cosmological constant model. We used the subscript/superscript ($0$)
to indicate that there is no particle production in this situation.
In addition, due to the universe expansion, the densities of
confined fluids, including radiation and etc., are diluted. This
provides a suitable situation in which the effects of this particle
production process overcomes those of the other fluids leading to
change the expansion phase. Bearing Eq.~(\ref{flbc1}) in mind, in
order to evaluate the effects of this pressure on the entropy of the
dominated fluid, we use Eq.~(\ref{entropy2n}) while $\rho=0$ to get
\begin{eqnarray}\label{en2}
\frac{dS}{dH}=\frac{\pi(1+\omega)}{H^3}(\frac{18\pi H^4(1+\omega)-16\pi QH}{3H^4(1+\omega)^2}-1),
\end{eqnarray}
where $H$ is the Hubble parameter when the effects of the particle
production is considered. In order to calculate the corresponding
thermal fluctuations due to this pressure, by using Eq.~(\ref{C1}),
we get
\begin{eqnarray}\label{S1}
CT^2=-\frac{3(1+3\omega^0)}{4},
\end{eqnarray}
leading to $dS^1=0$ for $\omega^0=\textmd{cons}$. Since this result
is independent of $p_C$, we can conclude that the gravitationally
induced particle production processes induce weak fluctuations into
the thermodynamic properties of the cosmos supported by a prefect
fluid with constant state parameter. It explains that why these
processes do not disturb the current thermodynamic equilibrium state
of the cosmos investigated in \cite{pavon3,LBC1}. In order to find
an expression for $S^2$, one can insert Eqs.~(\ref{en1})
and~(\ref{en2}) into~(\ref{totalentropy}) as
\begin{eqnarray}
\frac{dS^2}{dH_0}=\frac{dS}{dH}\frac{dH}{dH_0}-\frac{dS_0}{dH_0},
\end{eqnarray}
which yields
\begin{eqnarray}\label{last}
\frac{dS^2}{dH_0}=\frac{\pi(1+\omega)}{H^3}(\frac{18\pi H^2+6\Gamma }{3H^2(1+\omega)^2}-1)
\frac{dH}{dH_0}+\frac{\pi}{H_0^3}(1+3\omega^0),
\end{eqnarray}
for models obeying~(\ref{pre}). It is easy to show that
\begin{eqnarray}\label{hdotp0}
\dot{H}_0=-\frac{3}{2}H_0^2(1+\omega^0),
\end{eqnarray}
where $p_C=0$, and
\begin{eqnarray}\label{hdotp0}
\dot{H}=-\frac{3}{2}H^2(1+\omega)[1-\frac{\Gamma}{3H}],
\end{eqnarray}
while $p_C\neq0$ and obeys~(\ref{pre}). In deriving these equations
(Raychaudhuri equations) we have used Eqs.~(\ref{flbc1})
and~(\ref{emc2}). Now, since
\begin{eqnarray}
\frac{dH}{dH_0}=\frac{\dot{H}}{\dot{H}_0},
\end{eqnarray}
we get
\begin{eqnarray}\label{lastf}
\frac{dS^2}{dH_0}=\frac{\pi}{H_0^2}[\frac{(1+\omega)^2}{H(1+\omega^0)}(\frac{18\pi
H^2+6\Gamma }
{3H^2(1+\omega)^2}-1)(1-\frac{\Gamma}{3H})+\frac{(1+3\omega^0)}{H_0}].
\end{eqnarray}
Therefore, we find an expression for the fluctuations of the
dominated fluid entropy which are due to the gravitationally
induced particle production process. Finally, we should note that
since the thermal fluctuation theory allows to evaluate $S^2$ up
to the desired order \cite{landau}, one can use this equation to
find a general relation for the particle production rate $\Gamma$,
meeting Eq.~(\ref{pre}), in a compatible way with thermal
fluctuation theory. Indeed, our study is available whenever the
state parameter of the dominated prefect fluid be constant.
Loosely speaking, in cases with non-constant state parameter
$dS^1\neq0$ and one should use~(\ref{totalentropy1}), just the
same as the section $\textmd{III}$, to find an expression for the
thermal fluctuations of system up to the desired order. Such
situations are appearing in transition eras whiles, the dominated
prefect fluid is slowly replaced by the new one and therefore, the
state parameter $\omega$ is not a constant quantity. We should
note again that since one can consider a dominated prefect fluid
with constant state parameter $\omega$ during an expansion era,
between the two transition eras \cite{roos}, $dS^1=0$ is valid in
this situation meaning that the gravitationally induced particle
production process induces weak fluctuations into the system.
These fluctuations are of the second and higher orders which may
explain that why this model preserves its final thermodynamic
equilibrium state \cite{pavon3,LBC1}.
\section{Gravitationally particle production may modify the horizon entropy}
Here, we are going to study the entropy of apparent horizon of the
flat FRW universe, filled by a prefect fluid with state parameter
$\omega=\frac{p}{\rho}$, which is also affected by an additional
pressure $p_C$ due to the gravitationally induced particle
production process \cite{LBC}. In fact, since it is shown that the
DE candidate may add an additional term to the Bekenstein entropy of
the horizon \cite{cana,cana1,em}, we are going to investigate the
probable effects of the gravitationally induced particle production
process, as the DE candidate, on the entropy of the apparent
horizon. Moreover, since we work on the apparent horizon, we use the
Hayward-Kodama temperature which is in fact an obvious
generalization of the Black Holes temperature to the apparent
horizon of the FRW universe \cite{Bak,hel1,hel,GSL1,em}. From the
Friedmann equation~(\ref{flbc1}) and~(\ref{emc2}) we get
\begin{eqnarray}\label{flbc12}
2HdH=\frac{8\pi}{3}d\rho,
\end{eqnarray}
and
\begin{eqnarray}\label{emc22}
d\rho=-3H(\rho+p)dt-Qdt,
\end{eqnarray}
respectively. Combining these equations together to reach
\begin{eqnarray}\label{flbc123}
2HdH+\frac{8\pi}{3}Qdt=-8\pi H(\rho+p)dt.
\end{eqnarray}
Bearing the Hayward-Kodama temperature
($T=-\frac{H}{2\pi}(1+\frac{\dot{H}}{2H^2})$) in mind
\cite{hel,hel1,Bak,GSL1}, we can rewrite this equation as
\begin{eqnarray}\label{flbc123}
T(-2HdH-\frac{8\pi}{3}Qdt)=-4H^2(\rho+p)dt-2(\rho+p)dH.
\end{eqnarray}
Since $E=\rho V$ and $dV=-\frac{4\pi}{H^4}dH$, we get
$dE=-4\pi\rho H^{-4}dH-4\pi H^{-2}(\rho+p)dt$, whiles $E$ is the
associated energy due to the source $\rho$ confined to the
apparent horizon. By inserting these relations into
Eq.~(\ref{flbc123}), we obtain
\begin{eqnarray}\label{flbc4}
T(-2HdH-\frac{8\pi}{3}Qdt)=\frac{H^4}{\pi}dE+2(\rho-p)dH,
\end{eqnarray}
leading to
\begin{eqnarray}\label{flbc5}
T(-\frac{2\pi}{H^3}dH-\frac{8\pi^2}{3H^4}Qdt)=dE-WdV,
\end{eqnarray}
where $W=\frac{\rho-p}{2}$ is the work density
\cite{Cai2,Cai3,CaiKim,GSL1,hel,hel1}. Comparing this equation with
the first law of thermodynamics in cosmological setups
($TdS_A=dE-WdV$), one gets
\begin{eqnarray}\label{flbc6}
dS_A=-\frac{2\pi}{H^3}dH-\frac{8\pi^2}{3H^4}Qdt,
\end{eqnarray}
which yields
\begin{eqnarray}\label{flbc7}
S_A=\frac{A}{4}+\pi\int\frac{(1+\omega)\Gamma}{H^2}dt+B,
\end{eqnarray}
where $B$ is the integration constant, $Q=-\rho(1+\omega)\Gamma$,
and we have used the Friedmann equation~(\ref{flbc1}) to obtain the
second term of RHS of this equation. Using~(\ref{emc22}) together
with~(\ref{flbc1}), one can rewrite the second term of RHS of this
equation as
\begin{eqnarray}\label{flbc8}
\pi\int\frac{(1+\omega)\Gamma}{H^2}dt=\frac{3}{8}\int\frac{\Gamma
d\rho}{\rho^2(\Gamma-\sqrt{24\pi\rho})}.
\end{eqnarray}
Therefore, based on this equation, the DE candidate (gravitationally
induced particle production process), may add an additional term to
the horizon entropy which leads to modify the Bekenstein limit
($S_A=\frac{A}{4}$). Previously, it has been shown that the dynamic
candidates of DE, such as the ghost dark energy model and its
generalization, may modify the horizon entropy \cite{em,cana,cana1}.
Moreover, it is shown that any mutual interaction between the cosmos
sectors may lead to modify the Bekenstein limit of the horizon
entropy \cite{em}. Therefore, the additional term appearing in this
equation is in line with the mentioned attempts
\cite{cana,cana1,em}. It is also useful to note that in the absence
of a mutual interaction between the gravitational and baryonic
fields ($\Gamma=0$), the results of the previous works are recovered
as a desired expectance
\cite{Cai2,Cai3,CaiKim,GSL1,Bak,hel,hel1,Cai44,em}.
\section{Summary and concluding remarks}
In this study, we have investigated the thermodynamics of the
cosmoses with the initial and final de-Sitter spacetimes. With
this aim, we have showed that the state parameter of varying
vacuum, as the DE candidate with varying energy density, cannot be
equal to $-1$, whenever there is no mutual interaction between the
fluids supporting the FRW background except whiles the universe is
in the de-Sitter stages with $\dot{H}=0$ meaning that
$\dot{\rho}=0$. Thereinafter, we have derived an expression for
the entropy changes of the introduced DE model. In addition, we
have payed our attention to a universe in which the cosmos sectors
interact with each other, and we got an entropy change relation
due to this interaction. In continue, by focusing on the model
with $\omega_D=-1$, we have found its corresponding
thermodynamical interpretation. In fact, our study shows that the
mutual interaction between the varying vacuum and the other parts
of cosmos which disturbs the initial de-Sitter phase of the
universe (whiles $\dot{H}_0=0$) \cite{LBSV}, may also disturb the
thermodynamic equilibrium of initial de-Sitter spacetime (where
$dS_D^0=d^2S_D^0=0$), and leads to $dS_D\neq0$. Thereinafter,
since this interaction brings the ultimate de-Sitter spacetime
(where again $\dot{H}=\dot{H}_0=0$)~\cite{LBSV} which satisfies
again the thermodynamic equilibrium conditions \cite{pavon3},
$\dot{\rho}_D=0$ and thus $Q=0$. The latter means that
$dS_D=dS_D^0$ and thus, the thermodynamic equilibrium conditions
are marginally satisfied in the ultimate de-Sitter spacetime
($dS_D^0=d^2S_D^0=0$) which is in agreement with~\cite{pavon3}.
Therefore, it seems that this mutual interaction between the
initial and final de-Sitter eras may leave some thermal
fluctuations into the system. Finally, by bearing the thermal
fluctuation theory in mind, we got a relation between the thermal
fluctuations of the DE candidate and the mutual interaction
between the cosmos sectors. Moreover, we have pointed to the
models in which there is a particle production process induced by
the gravity. We showed that such process may be considered as an
interacting model whiles this interaction comes from the
gravitationally induced particle process. In fact, our resolution
shows that such particle production, coming from non-equilibrium
thermodynamic analysis, inspires a weak fluctuation to the
thermodynamical properties of the model, including the entropy and
temperature, and thus the spacetime features such as its apparent
horizon. We have found out a relation between the rate of the
particle production and these fluctuations which are second order
onwards. The latter may explain why in this model the
thermodynamic equilibrium state of cosmos does not change due to
the gravitationally induced particle production process
\cite{pavon3,LBC1}. In addition, we have tried to establish the
first law of thermodynamics on the apparent horizon and therefore,
we found that the gravitationally induced particle production
process, as the DE candidate, may add an additional term to the
Bekenstein entropy of the apparent horizon. The latter is in line
with the previous works claiming that the DE candidate and its
interaction with the other parts of cosmos may change the horizon
entropy \cite{cana,cana1,em}.

\acknowledgments{This work has been supported financially by
Research Institute for Astronomy \& Astrophysics of Maragha
(RIAAM).

\end{document}